\documentclass[reprint,aip,jcp,amsmath,amssymb,superscriptaddress,10pt]{revtex4-1}

\usepackage{graphicx}
\usepackage{graphics}
\usepackage{dcolumn}
\usepackage{bm}
\usepackage{color}
\usepackage{soul}
\usepackage{textcomp}
\usepackage{soul}
\usepackage{rotating}
\usepackage{setspace}
\usepackage[mathlines]{lineno}
\usepackage{textcomp}
\usepackage{mathcomp}

\newcommand{\valprom}[1]{\left<#1\right>}

\newcommand{\GFTyMA}{
Grupo de F{\'i}sica Te{\'o}rica y Matem{\'a}tica Aplicada, Instituto de F{\'i}sica, 
Facultad de Ciencias Exactas y Naturales, 
Universidad de Antioquia; Calle 70 No. 52-21, Medell\'in, Colombia.}
\newcommand{\GFAM}{
Grupo de F\'isica At\'omica y Molecular, Instituto de F\'{\i}sica,  
Facultad de Ciencias Exactas y Naturales, 
Universidad de Antioquia UdeA; Calle 70 No. 52-21, Medell\'in, Colombia.}

\usepackage{microtype}
\usepackage[breaklinks=true,hyperindex=true,
pdftitle={Environment-Assisted One-photon Coherent Phase Control},
colorlinks=true,pagebackref=false,citecolor=blue,plainpages=false,pdfpagelabels,
linkcolor=blue,urlcolor=blue]{hyperref}

\begin{document}


\title{
Influence of Non-Markovian Dynamics in Equilibrium Uncertainty-Relations
}

\title{
Influence of Non-Markovian Dynamics in Equilibrium Uncertainty-Relations
}
\author{Leonardo A. Pach\'on}
\affiliation{\GFTyMA}
\affiliation{\GFAM}
\author{Johan F. Triana}
\affiliation{\GFAM}
\author{David Zueco}
\affiliation{Instituto de Ciencia de Materiales de Arag\'on y Departamento de 
F\'isica de la Materia Condensada,
CSIC-Universidad de Zaragoza, Zaragoza E-50012, Spain.}
\affiliation{Fundaci\' on ARAID, Paseo Mar\'{\i}a Agust\'{\i}n 36, E-50004 Zaragoza, Spain.}
\author{Paul Brumer}
\affiliation{Chemical Physics Theory Group, Department of Chemistry and
Center for Quantum Information and Quantum Control,
University of Toronto, Toronto, Canada M5S 3H6}

\begin{abstract}
Contrary to the conventional wisdom that deviations from standard thermodynamics originate from
the strong coupling to the bath, it is shown that in quantum mechanics, these deviations originate
from the uncertainty principle and are supported by the non-Markovian character of the dynamics.
Specifically, it is shown that the lower bound of the dispersion of the total energy of the system, 
imposed by the uncertainty principle, is dominated by the bath power spectrum and therefore, 
quantum mechanics inhibits the system thermal-equilibrium-state from being described by the 
canonical Boltzmann's distribution.
We show that for a wide class of systems, systems interacting via central forces with 
pairwise-self-interacting environments, this general observation is in sharp contrast to the classical 
case, for which the thermal equilibrium distribution, irrespective of the interaction strength, is 
\emph{exactly} characterized by the canonical Boltzmann distribution and therefore, no dependence 
on the bath power spectrum is present.
We define an \emph{effective coupling} to the environment 
that depends on all  energy scales  in the system and reservoir interaction.
Sample computations in regimes predicted by this effective coupling are demonstrated.
For example, for the case of strong effective coupling,  deviations from standard thermodynamics 
are present  and, for the case of weak effective coupling, quantum features such as stationary 
entanglement are possible at high temperatures.
\end{abstract}

\date{\today}


\maketitle

\section{Introduction}
Thermodynamics was developed before the modern atomistic description of Nature was formulated.
Statistical mechanics  then led to an understanding of the laws of thermodynamics in terms
of a microscopic description, thus closing the gap between macroscopic and 
microscopic description.
Due to the interest in quantum technologies, there is a major ongoing effort  to develop a consistent
and well defined extension of thermodynamics to the quantum regime \cite{AN00,EG&08,CHT11}.
However, the majority of these theories are primarily based on Boltzmann's original ideas
and are  therefore  plagued by issues concerning irreversibility, the origin of the second law,
the relation between physics and information, the meaning of ergodicity, etc. (see, e.g., 
Refs.~\cite{GM03,GM04}).

Despite these issues, progress has been made. It is now well known  that, e.g., Onsager's 
regression hypothesis fails in the quantum realm \cite{Tal86,FO96} and that non-Markovian dynamics are relevant
in a variety of fields and applications, from foundations \cite{EG&08,DL13}, to nuclear
physics \cite{WS&13}, quantum metrology \cite{MBF11,HRP12,CHP12} and biological systems (see, e.g.,
\cite{PB12} and references therein).
It is also known that the thermal equilibrium state of a quantum system \emph{strongly} coupled
to a thermal bath deviates from the canonical Boltzmann distribution \cite{GWT84,GSI88,AN00,CTH09,SM07}.
This is also expected to occur in the classical case \cite{Jar04}.
Since both are incoherent thermal stationary situations, one would anticipate that the quantum
system is devoid of any coherence and hence, based on the decoherence program \cite{Zur03},
that both distributions should coincide.
However, one might also suggest that the entanglement between the system and the bath, which has
no classical counterpart, could introduce quantum-classical deviations \cite{AN00}.
Furthermore, the fact that extra deviations could be present even if the entanglement between
the system and the bath is zero \cite{HL09} or even in the weak field regime \cite{RO89,GRT00}
makes the situation even more intriguing.

To explore this issue in greater detail,
it is advantageous to find a situation where the classical and the quantum contributions
to the deviation from the Boltzmann distribution can be clearly isolated and examined.
Here we show that \emph{irrespective of the interaction strength}, 
there are no deviations from Boltzmann's
distribution for the generic case of a classical system 
interacting via central forces with a pairwise-self-interacting
environment.
Thus, if after quantum-mechanically treating the same case, deviations from the canonical
Boltzmann's distribution are present, then they are purely quantum in nature.
As shown below, deviations do appear not only for this systems but in general and, based 
on completely universal arguments, are shown to rely on the \emph{uncertainty principle} 
characteristic of quantum mechanics and the power spectrum of the environment.

Specifically, the uncertainty principle not only inhibits the system's thermal-equilibrium-state 
from being described by the canonical Boltzmann distribution, but for \emph{each system-bath 
interaction it also selects which thermal equilibrium states are physically accessible}.
This latter result, formulated here for the first time within the framework of quantum 
thermodynamics, constitutes the cornerstone of the theory of pointer states (the states 
which are robust against the presence of the environment) \cite{Zur03} and could have 
deep consequences for an  understanding of the thermalization of quantum systems.

We note that that Romero \& Oppenheim \cite{RO89}  as well as
Geva \textit{et al.}\cite{GRT00}previously addressed the issue, using a 
detailed master equation approach, of deviations from 
the quantum canonical state, even in the weak field regime. 
Here we display the quantum origin of those deviations for the large
family of physical systems presented above, and clearly display 
intriguing aspects of the strong and  weak coupling regimes.

\subsection{Classical Thermal-Equilibrium-State for a System 
Interacting via Central Forces with a Pairwise-Self-Interacting
Environment.}
\label{subsect:ThermalState}
The classical and quantum thermal state will turn out to be significantly 
different for the following general class of systems. We consider
a classical particle of mass $m$ with potential energy $U_{\mathrm{S}}(q)$ and Hamiltonian
%
$
H_{\mathrm{S}}(p,q) =
\frac{1}{2m}{p}^2 + U_{\mathrm{S}}(q),
$
%
and a bath of $\mathfrak{N}$ classical particles interacting 
via the central force potential 
$U^{\mathfrak{i},\mathfrak{j}}_{\mathrm{B}}(\mathfrak{q}_{\mathfrak{i}}-\mathfrak{q}_{\mathfrak{j}})$
and Hamiltonian
$
H_{\mathrm{B}}(\mathfrak{p},\mathfrak{q}) = 
\sum_{\mathfrak{j}}^{\mathfrak{N}}[ 
\frac{1}{2 \mathfrak{m}_{\mathfrak{j}} }{\mathfrak{p}}_{\mathfrak{j}}^2
+\sum_{\mathfrak{i},\mathfrak{j}}U^{\mathfrak{i},\mathfrak{j}}_{\mathrm{B}}
(\mathfrak{q}_{\mathfrak{i}}-\mathfrak{q}_{\mathfrak{j}})
].
$
The classical system interacts with this bath via the central force potential energy
$\mathcal{V}_{\mathfrak{j}} ({\mathfrak{q}}_{\mathfrak{j}} - {q})$, so that the total
Hamiltonian is given by
\begin{align}
\label{equ:cCFH}
\begin{split}
H &= H_{\mathrm{S}}(p,q) + H_{\mathrm{B}}(\mathfrak{p},\mathfrak{q})
+ \sum_{\mathfrak{j}}^{\mathfrak{N}}
\mathcal{V}_{\mathfrak{j}} ({\mathfrak{q}}_{\mathfrak{j}} - {q}).
\end{split}
\end{align}
In classical statistical mechanics, the thermal equilibrium distribution of the system S is defined by
\begin{equation}
\label{equ:cgenequilstate}
\begin{split}
\rho_\mathrm{S}(p,q) &= \frac{1}{Z}
\int
\prod_{\mathfrak{j}}^{\mathfrak{N}}\mathrm{d}\mathfrak{p}_{\mathfrak{j}} \mathrm{d}\mathfrak{q}_{\mathfrak{j}}
\exp\left[-H(p,q,{\mathfrak{p}}_{\mathfrak{j}},{\mathfrak{q}}_{\mathfrak{j}})
\beta \right],
 \end{split}
\end{equation}
where
$Z = \int \prod_{\mathfrak{j}}^{\mathfrak{N}}\mathrm{d}\mathfrak{p}_{\mathfrak{j}}
\mathrm{d}\mathfrak{q}_{\mathfrak{j}}
\int \mathrm{d}p \mathrm{d}q \exp(-H \beta)$
denotes the partition function of the total system, with $\beta = 1/k_{\mathrm{B}} T$
and $T$ being the temperature of the environment.
In reaching the equilibrium thermal state in Eq.~(\ref{equ:cgenequilstate}) [also in 
Eq.~(\ref{equ:qgenequilstate}) below], it is assumed that the central system S gets 
coupled at $t=-\infty$ to the initially thermalized bath B.
Therefore, the equilibrium state $\rho_\mathrm{S}(p,q)$ turns out to be an eigenmode 
of the full dissipative dynamics \cite{PYB13,PB13c,PB13}.

The integral over $\mathfrak{p}_{\mathfrak{j}}$ in Eq.~(\ref{equ:cgenequilstate}) trivially
cancels out with the corresponding contribution in $Z$ (see Appendix~\ref{app:ReducedClassState}
for details).
Due to the form of the dependence of $\mathcal{V}_{\mathfrak{j}}$ and 
$U^{\mathfrak{i},\mathfrak{j}}_{\mathrm{B}}$ on ${\mathfrak{q}}_{\mathfrak{i}}$, 
${\mathfrak{q}}_{\mathfrak{j}}$ and ${q}$, the integrals over $\{\mathfrak{q}_{\mathfrak{j}}\}$ 
can be appropriately manipulated, with the net result that they cancel out with the 
corresponding contribution in $Z$.
Thus,
\begin{equation}
\label{equ:equilstate}
\begin{split}
\rho_\mathrm{S}(p,q) &= {Z_\mathrm{S}}^{-1}
\exp\left[-H_{\mathrm{S}}(p,q)
 \beta \right],
 \end{split}
\end{equation}
where
$Z_{\mathrm{S}}  = \int \prod_{j}^{N} \mathrm{d}p_j \mathrm{d}q_j \exp(-H_{\mathrm{S}} \beta)$.
Hence, the thermal equilibrium distribution of a bounded particle in contact with a pairwise-self-interacting thermal bath
via central forces, \emph{irrespective of the coupling strength}, is exactly given by the
canonical Boltzmann distribution.
If the system S is composite, this result remains valid if each constituent of S is coupled 
to its own independent bath. 

This result is enlightening because, in the strong coupling regime, there is no apparent
physical reason why the equilibrium thermodynamic properties of a system are independent,
for a wide class of systems,
of both the nature of the bath to which it is coupled and of the functional form of the observables that mediate 
the interaction.
For this particular type of systems, the physical picture that emerges from this result is that 
\emph{in the long time regime}, any dissipative mechanism is equally effective in taking the 
system to thermal equilibrium.
In other words, dissipative dynamics can contract the classical phase-space volume with no
fundamental restriction and therefore, the resultant equilibrium state is independent of the
dissipative coupling and the rate at which equilibrium [Eq.~(\ref{equ:equilstate})] is reached.
This suggests that the concept of intrinsic and extensive thermodynamic variables \cite{GM03}
can be extended, in some cases, to the strong coupling regime.
By contrast, and as shown below,
in the quantum realm dissipative mechanisms are accompanied by decoherence effects and
are bath-nature and coupling-particularities sensitive \cite{PZ00, Zur03} and are then capable of
inducing a variety different thermal states.

\section{Quantum Thermal-Equilibrium-State}
The thermal-equilibrium-state of a classical system interacting via central forces with a 
pairwise-self-interacting environment was discussed above.
The quantum treatment discussed below extends beyond this particular type of system because it 
is performed on more general grounds.
In doing so, the total Hamiltonian in Eq.~(\ref{equ:cCFH}) is replaced here by the general 
expression 
$\hat{H}  = \hat{H}_{\mathrm{S}} + \hat{H}_{\mathrm{B}} + \hat{V}$.
Based on the general description given in \cite{CTH09,CHT11,GSI88}, one can easily extend
the classical definition in Eq.~(\ref{equ:cgenequilstate}) to the quantum regime, namely,
\begin{equation}
\label{equ:qgenequilstate}
\hat{\rho}_{\mathrm{S}} = \frac{1}{Z}\mathrm{tr}_{\mathrm{B}}
\exp\left\{-\left[ \hat{H}(\hat{p},\hat{q},{\hat{\mathfrak{p}}}_{\mathfrak{j}},{\hat{\mathfrak{q}}}_{\mathfrak{j}})
\right]\beta \right\}.
\end{equation}
The operator character of the various terms in this expression and their commutativity
relations prevent us from proceeding as we did in the classical case.
However, these very same commutativity relations allow the
immediate formulation of the following set of inequalities:
\begin{subequations}
\label{equ:nonconmbound}
\begin{align}
\label{equ:nonconmbound.a}
[\hat{H}_{\mathrm{S}}, \hat{V} ] \neq 0 \Rightarrow
\Delta \hat{H}_{\mathrm{S}} \Delta \hat{V} &\ge \frac{1}{2}
|\langle[ \hat{H}_{\mathrm{S}},\hat{V}] \rangle|,
\\
\label{equ:nonconmbound.b}
[\hat{V}, \hat{H}_{\mathrm{B}} ] \neq 0 \Rightarrow
\Delta \hat{V} \Delta \hat{H}_{\mathrm{B}} &\ge \frac{1}{2}
|\langle[\hat{V} ,\hat{H}_{\mathrm{B}}] \rangle|,
\end{align}
\end{subequations}
where
$\Delta \hat{O} = \sqrt{\langle \hat{O}^2 \rangle - \langle \hat{O} \rangle^2}$ denotes
the standard deviation of $\hat{O}$, with $\langle \hat{O} \rangle = \mathrm{tr}(\hat{O} \hat{\rho})$,
$\hat{\rho}$ being the thermal equilibrium state of the system S and the bath B.
Because of the thermal character of the density operator discussed here, Eqs.~(\ref{equ:nonconmbound})
incorporate classical as well as quantum uncertainty, however, the lower bound in both cases differs.
Equation ~(\ref{equ:nonconmbound}) applies to {\it any} system, bath,
and system-bath coupling Hamiltonians, but will prove particularly
noteworthy in the case of Eq.~(\ref{equ:cCFH}) where the classical
$\rho_{\rm S}$ is Boltzmann. 

For example, consider $[ \hat{H}_{\mathrm{S}},\hat{V}]= 0$. 
This implies a pure decohering interaction, which can be treated here in the framework 
of fluctuations without dissipation \cite{FO98}.
The equilibrium state is an incoherent mixture of system's eigenstates and is expected to be
well characterized by the canonical Boltzmann distribution \cite{DY&07}.
In this case $[ \hat{H}_{\mathrm{S}},\hat{V}]= 0$, so that $\Delta \hat{H}_{\mathrm{S}} \Delta \hat{V} \ge 0$,
meaning that the commutativity relation here may result in the minimum lower bound on
$\Delta \hat{H}_{\mathrm{S}} \Delta \hat{V}$.
Note that the same lowest limit may be obtained if, as in the classical case, the thermal equilibrium
state of the system $\hat{\rho}_{\mathrm{S}}$ is formally the canonical 
Boltzmann distribution $\hat{\rho}_{\mathrm{S}}^{\mathrm{can}}$.
Specifically, if
$\hat{\rho} = \hat{\rho}_{\mathrm{S}}^{\mathrm{can}}\otimes \mathrm{tr}_{\mathrm{S}}\hat{\rho}$, then
$|\langle[ \hat{H}_{\mathrm{S}},\hat{V}] \rangle| = \mathrm{tr}([ \hat{H}_{\mathrm{S}},\hat{V}] \hat{\rho})
=\mathrm{tr}([\hat{\rho}, \hat{H}_{\mathrm{S}}] \hat{V})=0$  since
$[\hat{\rho}_{\mathrm{S}}^{\mathrm{can}},\hat{H}_{\mathrm{S}}]=0$, giving
$\Delta \hat{H}_{\mathrm{S}} \Delta \hat{V} \ge 0$.
This is a consequence of the fact that the Boltzmann distribution is the zero-order-in-the-coupling
thermal equilibrium state.

The quantity $|\langle[ \hat{H}_{\mathrm{S}},\hat{V}] \rangle|$ is a measure of the quantum
correlations between the system and the bath and Eqs.~(\ref{equ:nonconmbound})
dictate the lower bound of
$\Delta \hat{H}_{\mathrm{S}}\Delta \hat{V}$.
This lower bound is different for each interaction since  each particular form of
$\hat{V}$ imposes a different commutation relation.
This last statement is precisely what allows, for example,  for a connection between 
the theory of pointer states and quantum thermodynamics.
If $\hat{B}$ and $\hat{S}$ denote observables of the bath and the system, respectively,
the dependence on the type of interaction can be obtained by calculation. For example, if the 
interaction term is chosen to be $\hat{V} = \hat{B}\otimes \hat{S}$ and the commutator 
$[ \hat{H}_{\mathrm{S}},\hat{V}]$ is calculated to second order in $\hat{V}\beta$, this 
leads to (see Appendix~\ref{app:FourthOrder})
\begin{equation}
\label{equ:thermodynbound}
|\langle[ \hat{H}_{\mathrm{S}},\hat{V}] \rangle| \propto
\mathrm{tr}_{\mathrm{S}}
\left\{
[\hat{H}_{\mathrm{S}}, \hat{S}]\mathrm{e}^{-\hat{H}_{\mathrm{S}}\beta}
\int\limits _0^{\hbar \beta} \mathrm{d}\sigma \hat{S}(-\mathrm{i}\sigma) K(\sigma)
\right\},
\end{equation}
where
$\hbar K(\sigma) = \langle \hat{B}(-\mathrm{i}\sigma)\hat{B}(0) \rangle_{\mathrm{B}}$
denotes the two-time correlation of the bath operators. 

Note that as long as second order perturbation theory is valid, Eq.~(\ref{equ:thermodynbound})
holds for any $\hat{S}$ and $\hat{B}$ and can be straightforwardly generalized to
the case of $\hat{V}= \sum_{\alpha}\hat{B}_{\alpha}\otimes \hat{S}_{\alpha}$.

A similar series-expansion analysis gives
$
\hat{\rho}_{\mathrm{S}} \propto
\mathrm{e}^{-\hat{H}_{\mathrm{S}}\beta}[ 1 + \hbar^{-1}
\int_0^{\hbar \beta} \mathrm{d}\sigma
\int_0^{\sigma} \mathrm{d}\sigma'
 \hat{S}(-\mathrm{i}\sigma)  \hat{S}(-\mathrm{i}\sigma') K(\sigma-\sigma')
].
$
Thus, for non-singular $K(\sigma)$, the thermal equilibrium state $\hat{\rho}_{\mathrm{S}}$ 
formally approaches the canonical Boltzmann distribution only when $\hbar \beta \rightarrow 0$.
Since each interaction is characterized by a given observable $\hat{S}$ and a given two-point
correlation function $K(\sigma)$, \emph{the general bounds in Eqs.~(\ref{equ:nonconmbound}, 
\ref{equ:thermodynbound})  predict different thermal equilibrium states for each type of interaction.}
Below, it shown that the bath spectrum is also related to the lower bound and thus, 
Eq.~(\ref{equ:nonconmbound}) also allows for a clear connection to other fundamental features 
such as the failure of the Onsager's hypothesis in the quantum regime \cite{Tal86,FO96}.

\section{Influence of the Spectrum of the Bath: Non-Markovian Character at Thermal Equilibrium}
Although the set of inequalities (\ref{equ:nonconmbound}, \ref{equ:thermodynbound}) are general, 
it is not possible to infer the role that, e.g., the spectrum of the bath plays in establishing the 
thermodynamic bounds above.
To provide a concrete expression for the lower bound in Eq.~(\ref{equ:nonconmbound.a}),
we focus on the case of 
$\hat{V} = \sum_{\mathfrak{j}}^{{\mathfrak{N}}} \hat{\mathcal{V}}_{\mathfrak{j}}(\hat{\mathfrak{q}}_{\mathfrak{j}} - \hat{q})$, set 
the interaction between the bath particles to zero, i.e., $U^{\mathfrak{i},\mathfrak{j}}_{\mathrm{B}}=0$ 
and consider the second order picture of the system-bath central force interaction, i.e.,
$\hat{V} \approx \sum^{\mathfrak{N}}_{\mathfrak{j}} \frac{1}{2}\mathfrak{m}_{\mathfrak{j}} \omega_{\mathfrak{j}}^2
\left(\hat{q}_{\mathfrak{j}} - \hat{q}\right)^2$, which yields to the well-known Ullersma-Caldeira-Leggett
model \cite{Ull66,CL83}.
After expanding $\hat{V}$,
it is possible to redefine
$\hat{H}$ in Eq.~(\ref{equ:cCFH}) as
$\hat{H}  = \hat{H}_{\mathrm{S}}' + \hat{H}_{\mathrm{B}}' + \hat{V}_{\mathrm{SB}}$ with
$\hat{V}_{\mathrm{SB}}= \hat{B}\otimes \hat{S}$.
Here $\hat{B}=\sum_{\mathfrak{j}}^{\mathfrak{N}} \mathfrak{m}_{\mathfrak{j}} \omega_{\mathfrak{j}}^2
\hat{q}_{\mathfrak{j}}$ and $\hat{S} = \hat{q}$, which act in the Hilbert space of the bath and the system,
respectively.
%
$\hbar K(\sigma) = \langle \hat{B}(-\mathrm{i}\sigma)\hat{B}(0) \rangle_{\mathrm{B}}$
given by  \cite{Ing02}
\begin{equation}
K(\sigma)
= \int \frac{\mathrm{d}\omega}{\pi}  J(\omega) \cosh\left(\tiny{\frac{1}{2}}\hbar \beta \omega - \mathrm{i}\sigma\right)
/\sinh\left(\tiny{\frac{1}{2}}\hbar \beta \omega \right),
\end{equation}
with $J(\omega) = \pi \sum_{\mathfrak{j}}^{\infty} \tiny{\frac{1}{2}}\mathfrak{m}_\mathfrak{j}
\omega_\mathfrak{j}^3\delta(\omega-\omega_\mathfrak{j}),
$ the spectral density of the bath.

The main feature of the quantum thermodynamic bound in
Eq.~(\ref{equ:thermodynbound}) is the presence of the power spectrum of the bath
$I(\omega,T) = \hbar J(\omega)\coth(\tiny{\frac{1}{2}}\hbar \beta \omega)$, which for bare Ohmic
dissipation, $J(\omega)=m\gamma \omega$, at high temperatures, $\hbar \beta \rightarrow 0$, 
is flat ($\omega$-independent) $I(\omega,T) \approx 2 m \gamma k_{\mathrm{B}} T$.
This is also true for higher orders in the serie expansion (see Appendix~\ref{app:FourthOrder}).
In this high temperature limit, and for non-singular $K(\sigma)$'s, the upper limit of the integral in 
Eq.~(\ref{equ:thermodynbound}) vanishes, leading to the vanishing of the commutator, even if 
$[\hat{H}_{\mathrm{S}}, \hat{S}]\neq 0$.
In other words, in the high temperature limit quantum correlations between the bath and 
the system disappear and the thermal equilibrium state is described by the canonical 
distribution, irrespective of the coupling strength or the functional form of the spectral 
density $J(\omega)$. 

For out-of-equilibrium quantum dynamics, the low temperature condition, finite $\hbar\beta$,
is associated with non-Markovian dynamics \cite{Ing02,HJ07}.
Since at fixed $T$, this non-Markovian character can be modified by the functional form of
the spectral density \cite{PB14a}, Eq.~(\ref{equ:thermodynbound}) makes clear that the 
equilibrium system properties depend on the non-Markovian character. 
This means that the quantum equilibrium statistical properties of a system experiencing 
Markovian dynamics (flat spectrum) are expected to differ from those of the same system 
experiencing non-Markovian dynamics (non-flat spectrum), which is in sharp contrast to
the classical case (see Eq.~\ref{equ:equilstate}).
This can be clearly understood in terms of the different thermodynamic lower bounds
resulting from either Markovian or non-Markovian interactions [see Eq.~(\ref{equ:thermodynbound})].

We make a connection with the failure of Onsager's regression hypothesis 
in quantum mechanics \cite{Tal86,FO96}, an hypothesis which is valid under Markovian dynamics \cite{Tal86,FO96} 
and when correlations between the bath and the system are negligible at equilibrium (in general, 
at any time) \cite{Swa81}.
Since formal Markovian dynamics can only be achieved for flat spectra \cite{PB14a} 
and for well behaved $J(\omega)$ with sufficient effective weak coupling (high temperature regime, 
see below) \cite{Kri13}, 
these two 
conditions can be seen as a single one when formulated in terms of Eq.~(\ref{equ:thermodynbound}).
Specifically, Markovian dynamics imply $|\langle[ \hat{H}_{\mathrm{S}},\hat{V}] \rangle| \rightarrow 0$
and, hence, the system-bath correlations vanish. 
This implies that Onsager's regression hypothesis, as well as the Boltzmann distribution, pertains 
exclusively to the classical realm.
Note that this statement is general and valid as long as second order perturbation theory 
is valid.

To provide some insight into the magnitude and consequences of the fundamental limit
derived above, an effective coupling to the bath is introduced below and two complementary regimes 
are analyzed.
Specifically, (i) an effective strong coupling characterized by deviations from standard 
thermodynamics, and (ii) an effective weak coupling that is shown below to allow for the 
survival of entanglement even between two oscillators in thermal equilibrium at high temperatures.

\subsection{Effective Coupling to the Bath.}
For the Ullersma-Caldeira-Leggett model, a standard calculation \cite{Ing02},
after removing a local contribution in the correlation function, yields
$
K(\tau)=
\frac{2m}{\hbar\beta}\frac{\mathrm{d}}{{\mathrm d}\tau}\sum_{l=1}^{\infty}
\tilde{\gamma}(|\nu_{l}|)\sin(|\nu_{l}|\tau),
$
where $\nu_{l}= l \Omega$, with $\Omega = 2\pi /\hbar\beta$ and $\tilde{\gamma}(z)$ 
defines an effective coupling to the bath.
Note that $\tilde{\gamma}(|\nu_{l}|)$ contains all the information about the 
correlations of the bath operators and therefore defines the influence of the 
bath on the system at thermal equilibrium.

For the subsequent discussion we adopt the most commonly used spectral density: 
the regularized Drude model with a high frequency cutoff $\omega_{\mathrm{D}}$,
$
J(\omega)   =  m_0 \gamma\omega\,
\omega_{\mathrm{D}}^2/\left(\omega^2 + \omega_{\mathrm{D}}^2\right),
$
where $\gamma$ is the standard coupling strength constant to the thermal bath
and $\omega_{\mathrm{D}}$ dictates the degree of non-Markovian dynamics.
For a discussion on the experimental reconstruction of the spectral density of open
quantum systems see Ref.~\cite{PB14}.
For the spectral density at hand,
\begin{align}
\label{equ:gammaD}
\tilde{\gamma}(|\nu_{l}|)=
\frac{\gamma}{ 1+ | l | \Omega / \omega_{\mathrm{D}}}.
\end{align}
Below we analyze the effective strong coupling, $\Omega/ \omega_{{\mathrm{D}}} \ll 1$,
and the effective weak coupling, $\Omega/ \omega_{{\mathrm{D}}} \gg 1$, regimes.

\textit{Strong Effective Coupling Regime}\textemdash
To quantify the consequences of non-flat spectra in this regime,   consider as the system a harmonic
oscillator of mass $m_0$ and frequency $\omega_0$ coupled to a thermal bath
 \cite{GWT84,GSI88}. See Ref.~\cite{PID10} for a phase-space description of the dynamics and 
 equilibrium  characteristics of this model.
In particular, we are interested in quantifying:   (i) the generation of squeezing in the
thermal equilibrium state, (ii) the deviation from the canonical partition function
$Z_{\mathrm{can}} = \mathrm{tr}_{\mathrm{S}}\mathrm{e}^{\hat{H}_{\mathrm{S}}\beta}$
and (iii) the deviation from the canonical von-Neumann entropy
$S_{\mathrm{can}}  = \mathrm{tr}_{\mathrm{S}}\left[\hat{\rho}_{\mathrm{can}}
\mathrm{ln}(\hat{\rho}_{\mathrm{can}})\right]$.

For this case the momentum and position variances are given by \cite{GWT84,GSI88}
$
\langle p^2 \rangle = m_0^2 \omega_0^2 \langle q^2 \rangle + \Delta,
$
where
$
\langle q^2 \rangle = \langle q^2_{\mathrm{cl}} \rangle +
\frac{2}{m_0 \beta}\sum_{n=1}^{\infty} \left[\omega_0^2 + \nu_n^2 +  \tilde{\gamma}(|\nu_n|)
|\nu_n|\right]^{-1}
$
and the squeezing parameter
$\Delta = - 2m_0 \gamma\beta^{-1} \partial \ln Z' /\partial \gamma$ with
$Z'=\frac{1}{\hbar \beta\omega}
\prod_{n=1}^{\infty} |\nu_n|\left[\omega_0^2 + \nu_n^2 +  \tilde{\gamma}(|\nu_n|)
|\nu_n|\right]^{-1}.
$
We recall that for this model, the classical theory predicts
$\langle p^2_{\mathrm{cl}} \rangle = m_0^2 \omega_0^2 \langle q^2_{\mathrm{cl}} \rangle$
and $\langle q^2_{\mathrm{cl}} \rangle = k_{\mathrm{B}}T/m_0 \omega_0^2$, so that
$\Delta_{\mathrm{cl}}=0$.
For the effective weak coupling regime $\Omega / \omega_{\mathrm{D}}\gg 1$,
disgarding terms of the order $\omega_0/\omega_{\mathrm{D}}$ and
$\gamma/\omega_{\mathrm{D}}$ gives 
$
\Delta \approx \pi \hbar \gamma m \omega_{\mathrm{D}}/ 6 \Omega
$
\cite{GWT84}.
Thus $\Delta$ vanishes at high temperatures, and the classical unsqueezed state
is recovered. However,
for the strong coupling regime $\Omega / \omega_{\mathrm{D}}\ll1$,
$
\Delta \approx \hbar \gamma m\,
 \mathrm{ln} \left(2\pi  \omega_{\mathrm{D}}/ \Omega\right)
$
\cite{GWT84}, meaning that the deviation from the canonical state translates into squeezing
of the equilibrium state.
This feature may be of relevance toward the generation of non-classical states, e.g., in
nano-mechanical resonators.

\begin{figure*}[ht]
\includegraphics[width=\linewidth]{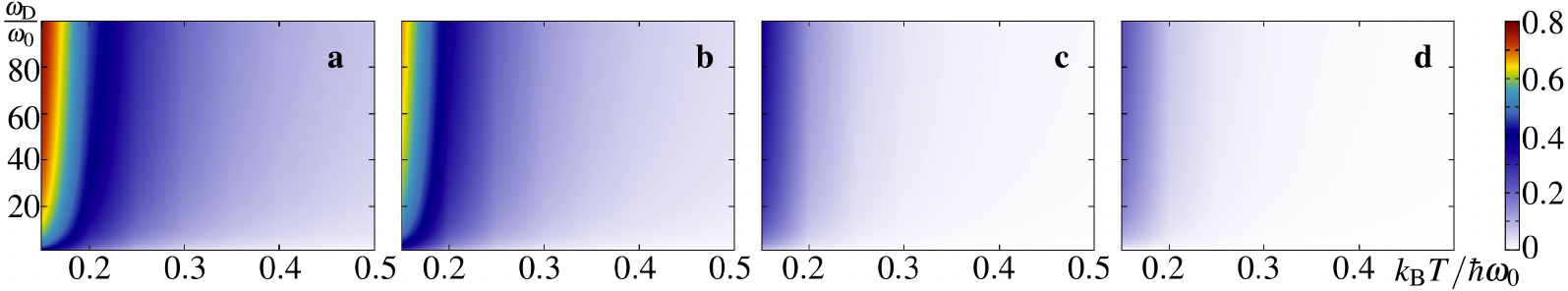}
\vspace{-0.5cm}
\caption{  \small $\log(Z/Z_{\mathrm{can}})$ for a harmonic oscillator as a function of the ratios
$k_{\mathrm{B}}T/\hbar \omega_0$ and $\omega_{\mathrm{D}}/\omega_0$.
We compare the partition function for $\gamma = 0.1\omega_0$ (a), $\gamma = 0.05\omega_0$ (b),
$\gamma = 0.01\omega_0$ (c) and $\gamma = 0.005\omega_0$ (d).
}
\label{fig:Entropy}
\end{figure*}
Deviations from the canonical result are also evident in the partition function $Z$.
Figure ~\ref{fig:Entropy} shows the logarithm of the ratio of $Z$ to the canonical partition
function $Z_{\mathrm{can}}$ as a function of the dimensionless parameters
$k_{\mathrm{B}}T/\hbar \omega_0$ and $\omega_{\mathrm{D}}/\omega_0$ for
(from left to right) $\gamma=0.1\omega_0$, $\gamma=0.05\omega_0$, $\gamma=0.01\omega_0$ and
$\gamma=0.005\omega_0$.
Deviations are observed at low temperatures and for high cutoff frequencies (i.e., in the
effective strong coupling regime).
In the opposite limit, regardless of the coupling parameter $\gamma$, both calculated
partition functions show the same behavior, as expected from the discussion above.
For the von Neumann entropy
$
S  = \mathrm{tr}_{\mathrm{S}}\left[\hat{\rho}_{\mathrm{S}}
\mathrm{ln}(\hat{\rho}_{\mathrm{S}})\right],
$
%
the behaviour of the ratio $\log(S/S_{\mathrm{can}})$ is  essentially the same as the one
described for the partition function ratio in Fig.~\ref{fig:Entropy}, and is not shown here.

\begin{figure*}
\includegraphics[width=1.03\linewidth]{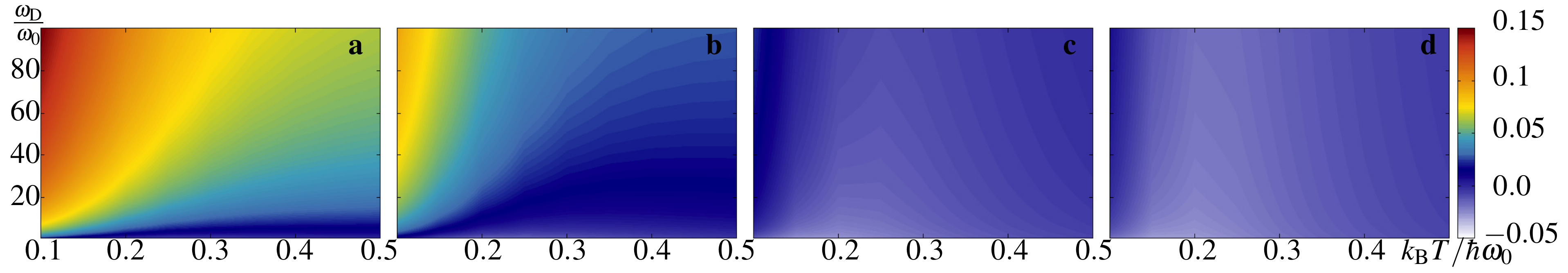}
\vspace{-0.5cm}
\caption{  \small Difference between the bounds quantum and classical 
in Eqs.~(\ref{equ:nonconmbound}) for a harmonic oscillator as a function of the ratio
$\omega_{\mathrm{D}}/\omega_0$.
We show cuts for different $k_{\mathrm{B}}T/\hbar \omega_0$ for $\gamma = 0.1\omega_0$ (a), $\gamma = 0.05\omega_0$ (b), $\gamma = 0.01\omega_0$ (c) and $\gamma = 0.005\omega_0$ (d).
}
\label{fig:boundd}
\end{figure*}

Since 
$\tilde{\gamma}(z) = \frac{1}{m_0}\int_0^{\infty}\frac{\mathrm{d}\omega}{\pi} \frac{J(\omega)}{\omega}
\frac{2z}{\omega^2+z^2}$,
each spectral densities defines a particular functional form of the effective 
coupling and therefore, of the thermal equilibrium properties.
Hence, as long as $\hbar\beta$ remains finite, different spectral densities lead 
to different thermal equilibrium states.
For the case of pair-wise central forces interactions, this can be considered as a quantum effect.

For this model, the bound in Eq.~(\ref{equ:nonconmbound.a}) is given by 
$\Delta\hat{H}_{\mathrm{S}}\Delta\hat{q}=
\sqrt{\frac{1}{2m_0^2} \langle \hat{p}^2\rangle^2 + \frac{1}{2} m_0^2 \omega_0^4 
\langle \hat{q}^2\rangle^2 - \frac{1}{4}\hbar^2\omega_0^2}\sqrt{\langle\hat{q}^{2}\rangle}$.
Contrary to the classical case where the bound is only determined by the temperature,
$(\Delta\hat{H}_{\mathrm{S}}\Delta\hat{q})_{\mathrm{cl}}=
k_{\mathrm{B}}T\sqrt{k_{\mathrm{B}}T/m\omega^{2}}$, the influence of the bath spectrum 
is clear in the quantum calculation of $\Delta\hat{H}_{\mathrm{S}}\Delta\hat{q}$. 
Figure~\ref{fig:boundd} depicts the difference 
$\Delta\hat{H}_{\mathrm{S}}\Delta\hat{q} - (\Delta\hat{H}_{\mathrm{S}}\Delta\hat{q})_\mathrm{cl}$ 
between the quantum and the classical bound for the $\gamma$ kernel in Eq.~(\ref{equ:gammaD}) 
as a function of the ratio $\omega_{\mathrm{D}}/\omega_0$ for different values of the ratio 
$k_{\mathrm{B}} T/\hbar \omega_0$.
The influence of the non-Markovian character is clear.


\textit{Weak Effective Coupling Regime}\textemdash
As an example, consider the survival of
entanglement at thermal equilibrium between two identical harmonic oscillators with masses
$m_0$ and frequencies $\omega_0$ linearly coupled with coupling constant $c_0$.
The Hamiltonian is given by
\begin{equation}
\hat{H}= \hat{H}_{\mathrm{S}} +
\sum_{\mathfrak{j},\alpha}^{\mathfrak{N},2} \left[\frac{\hat{p}_{\mathfrak{j},\alpha}^2}{2m_{\mathfrak{j}}}
+ \frac{m_{\mathfrak{j}} \omega_{\mathfrak{j}}^2}{2}
\left(\hat{q}_{\mathfrak{j},\alpha} - \hat{q}_{\alpha}\right)^2\right],
\label{equ:HamilEntang}
\end{equation}
with $\alpha=\{1,2\}$ and
$
\hat{H}_{\mathrm{S}} = \frac{1}{2m_0}(\hat{p}_{1}^{2}+\hat{p}_{2}^{2})
+\frac{1}{2}m_0 \omega_0^{2}( \hat{q}_{1}^2 + \hat{q}_{2}^2)
- c_0 q_1 q_2
$  \cite{NA02}.
The introduction of independent baths for each oscillator ensures that no deviations from
Boltzmann's distribution are present in the classical case.
This can be verified directly from the multi-particle-system generalization of Eq.~(\ref{equ:cCFH}).

\begin{figure*}[ht]
\includegraphics[width=\linewidth]{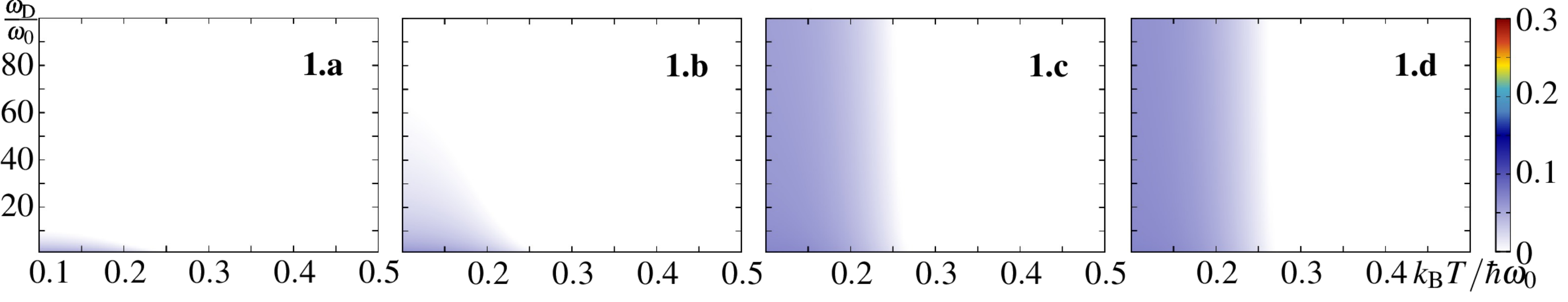}
\\
\includegraphics[width=\linewidth]{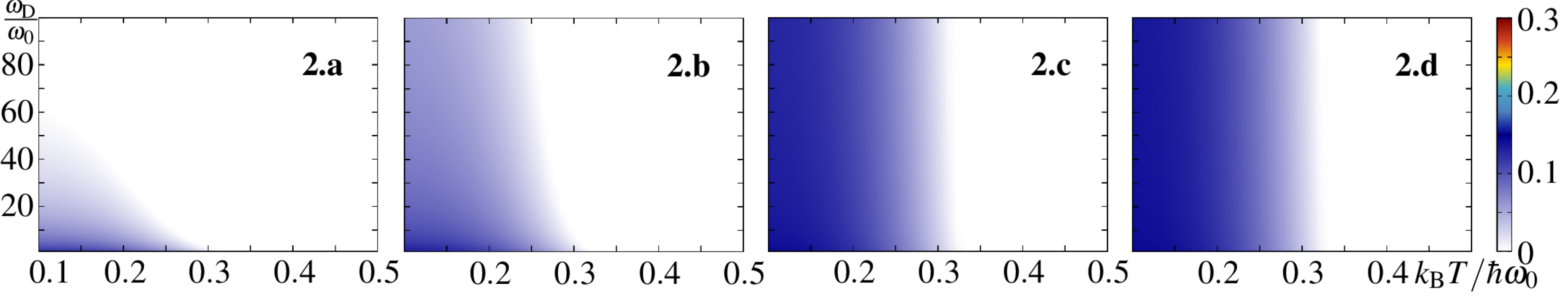}
\\
\includegraphics[width=\linewidth]{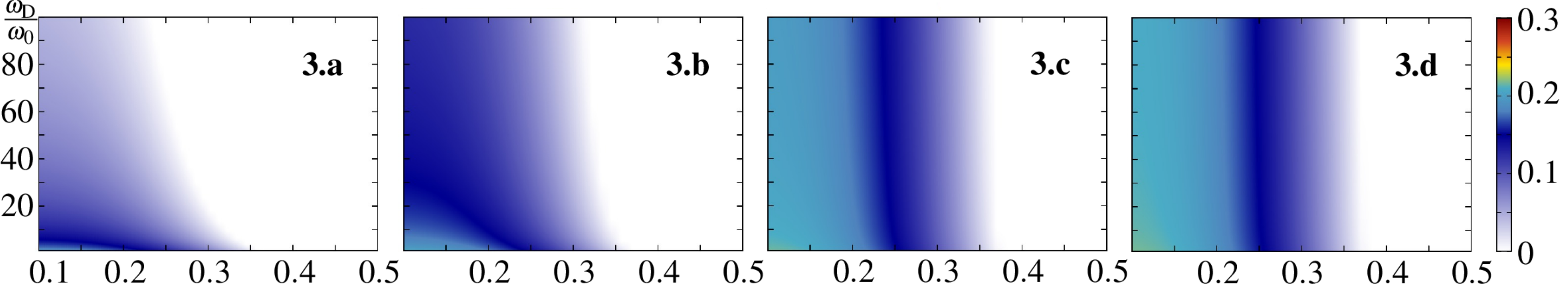}
\\
\includegraphics[width=\linewidth]{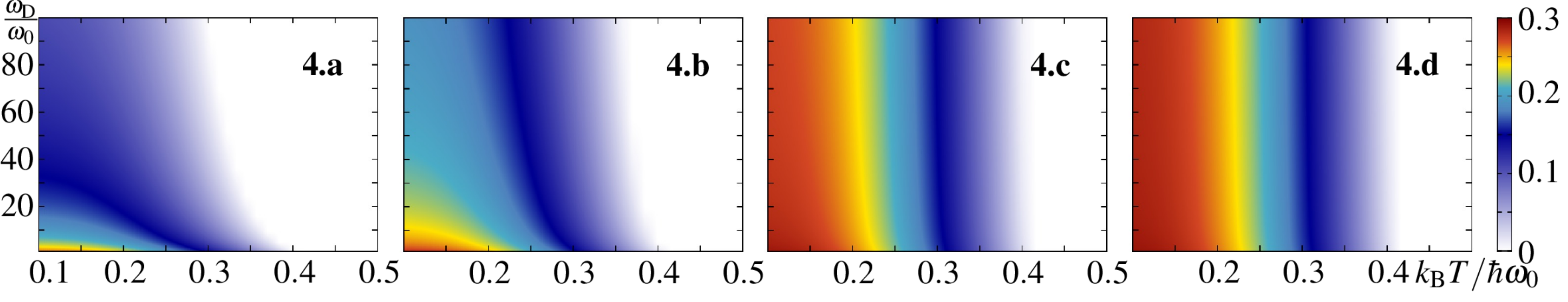}
\caption{\small Logarithmic Negativity in the presence of non-Markovian interactions
for $c_0 = 0.05 m_0 \omega_0^2$ first row,
$c_0 = 0.10 m_0 \omega_0^2$ second row,
$c_0 = 0.15 m_0 \omega_0^2$ third row and
$c_0 = 0.20 m_0 \omega_0^2$ fourth row.
Parameters are $\gamma = 0.1\omega_0$ (a), $\gamma = 0.05\omega_0$ (b),
$\gamma = 0.01\omega_0$ (c) and $\gamma = 0.005\omega_0$ (d) as a function of the dimensionless
parameters $k_{\mathrm{B}}T/\hbar \omega_0$ and $\omega_{\mathrm{D}}/\omega_0$.
}
\label{fig:EntanC}
\end{figure*}
At equilibrium, the entanglement between the two harmonic oscillators
can survive only when $k_{\mathrm{B}}T/\hbar \omega_0 \ll1$ \cite{GPZ10}.
However, this limit only applies in the Markovian regime and $\gamma\rightarrow 0$.
Thus, based on the discussion above, and supported by the recent observation that non-Markovian
dynamics assists entanglement in the longtime limit \cite{HRP12}, this
limit needs to be refined in order to account for the non-Markovian character of the interaction
and the finite value of $\gamma$ \cite{EP15}.

Due to the continuous-variable and Gaussian character of the systems, entanglement is quantified 
below by means of the logarithmic negativity $E_{\mathrm{N}}$.
This measure makes use of the positivity properties of the covariance matrix $\sigma$ of the 
Gaussian state that defines the full sate of the system\cite{GPZ10,EP15,TPE15} whose matrix
elements are defined
$\sigma_{ij}=\frac{1}{2}\valprom{\hat{r}_i \hat{r}_j+\hat{r}_j \hat{r}_i} - 
\valprom{\hat{r}_i} \valprom{\hat{r}_j}$,
where $\hat{\boldsymbol{r}}=(\hat{q}_1,\hat{q}_2,\hat{p}_1,\hat{p}_2)$ and 
$\hat{q}_1$, $\hat{q}_2$, $\hat{p}_1$, $\hat{p}_2$ are the position and 
momentum operators of the oscillators in the system of interest $\hat{H}_{\mathrm{S}}$
in Eq.~(\ref{equ:HamilEntang}).

$E_{\mathrm{N}}$ is defined in terms of the eigenvalues $l_i$'s of 
$-\mathrm{i}\Sigma\sigma$, with $\Sigma=\left(\begin{array}{cc}
0&\mathsf{I}_2\\
-\mathsf{I}_2&0
\end{array}\right)
$
the symplectic matrix and $\mathsf{I}_2$ the identity matrix of dimension 2.
Specifically \cite{VW02},
\begin{equation}
E_{\mathrm{N}}=-\frac{1}{2}\sum_{i=1}^4\log_2[\min(1,2|l_i|)].
\end{equation} 
The logarithmic negativity of the system is zero for separable states, 
$\hat{\rho}_{\mathrm{S}}=\sum_i p_i \hat{\rho}_1^{(i)}\otimes \hat{\rho}_2^{(i)}$,
but is unbounded from above since for continuous variable systems, the maximally 
entangled EPR wave-function has $E_{\mathrm{N}}\to\infty$.

For different values of the coupling constant $c_0$, Fig.~\ref{fig:EntanC} shows the 
logarithmic negativity for a variety of values of the damping constant $\gamma$ as a 
function of the dimensionless ratios $k_{\mathrm{B}}T/\hbar \omega_0$ and
$\omega_{\mathrm{D}}/\omega_0$.
As expected, (i) the more coupled the oscillators are, the higher the temperature and
the damping rate at which entanglement can survive at equilibrium, and (ii)
the smaller the damping rate (the more isolated the system) is, the higher the temperature
at which entanglement can be maintained.
The new feature here is that the more non-Markovian the interaction is, the higher the
temperature and the damping rate at which entanglement can be maintained \emph{at equilibrium}.
This behaviour is also present at out-of-equilibrium \cite{EP15} and can be easily interpreted in
terms of the effective coupling introduced above.

\section{Discussion.}
We have shown the role of the Heisenberg uncertainty principle [Eq.~(\ref{equ:thermodynbound})] 
in preventing quantum systems from relaxing to  the Gibbs state that is dictated by the system 
Hamiltonian only.  
The Gibbs state is only recovered in the classical-high $T$ limit ($\hbar \beta \rightarrow 0$).  
The implications at low-$T$ for quantum thermodynamics are crucial, such as the failure for the 
Onsager hypothesis, or the difficulty in defining the specific heat \cite{IHT09},
and the temperature definition \cite{FGA12}.
Specifically,  the  high  temperature  regime defined by $\hbar \beta  \rightarrow  0$,  modified  
by  an appropriate effective coupling,  emerges  as  the main condition for the vanishing of
these deviations.

For the wide class of classical systems discussed above (for which the 
thermal equilibrium state exactly corresponds to the Boltzmann distribution), the dependence 
of thermal equilibrium state on the spectral density is clearly a pure quantum effect.
This feature can be explored as a quantum resources, e.g., in the one-photon phase control of
biochemical and biophysical systems \cite{PB12,PB13,PB13c} or in understanding the coherent extent
of excitation with incoherent light in biological systems \cite{PB13,PBB17}.

Although  the  examples presented are specific to  the  second   order  approximation  
of  the  interaction potential,  the  general  picture  provided here remains valid, albeit more involved,
in non-linear cases and with non-Gaussian statistics.
The  results  presented  here clarify the role of non-Markovian dynamics  and its relevance at 
thermal equilibrium, and provide physical insights into how non-Markovian interactions
protect quantum features such as entanglement.
They  may  shed  light  on the role of non-Markovian dynamics in 
the derivation of fundamental limits  in  areas  such  as  quantum metrology \cite{CHP12},
quantum speed limits \cite{DL13} and cooling of nanomechanical 
resonators \cite{TPE15}.

Finally, if $\mathcal{D}_{\mathrm{S}}$ denotes the dimension of the Hilbert space of the system and 
$\mathcal{D}_{\mathrm{B}}^{\mathrm{eff}}$ denotes an abstract effective dimension of the Hilbert 
space of the bath, it is interesting to note that in the context of canonical typicality \cite{PSW06,GL&06},
the trace distance 
$\langle D(\hat{\rho}_{\mathrm{S}}, \hat{\rho}^{\mathrm{can}})\rangle = (1/2)\mathrm{tr} 
\sqrt{(\hat{\rho}_{\mathrm{S}} -  \hat{\rho}^{\mathrm{can}})^\dag
(\hat{\rho}_{\mathrm{S}} - \hat{\rho}^{\mathrm{can}})}$ 
between 
the general equilibrium state $\hat{\rho}_{\mathrm{S}}$ and the canonical distribution 
$\hat{\rho}^{\mathrm{can}}$ is bounded from above by the ratio 
$\frac{1}{2} \sqrt{\mathcal{D}_{\mathrm{S}}/\mathcal{D}_{\mathrm{B}}^{\mathrm{eff}}}$.
Formally, when the spectral density is introduced, the limit $\mathcal{D}_{\mathrm{B}} \rightarrow \infty$
is implicit. 
Thus, if $\mathcal{D}_{\mathrm{B}}^{\mathrm{eff}} = \mathcal{D}_{\mathrm{B}} $, contrary to the results 
presented here, no deviations from the canonical state are expected.
However, based on the results above, the trace distance is given here by
$D(\hat{\rho}_{\mathrm{S}}, \hat{\rho}^{\mathrm{can}}) \propto
\hbar^{-1} \mathrm{tr}_{\mathrm{S}}
\int_0^{\hbar \beta} \mathrm{d}\sigma
\int_0^{\sigma} \mathrm{d}\sigma'
 \hat{S}(-\mathrm{i}\sigma)  \hat{S}(-\mathrm{i}\sigma') K(\sigma-\sigma').
 $
This situation suggests that the abstract effective dimension $\mathcal{D}_{\mathrm{B}}^{\mathrm{eff}}$
must be a function of the ratio $\hbar/k_{\mathrm{B}}T$, the power noise of the environment $I(\omega)$ 
and of the observable that couples the system to the environment.
Deriving the specific functional relationship $\mathcal{D}_{\mathrm{B}}^{\mathrm{eff}}[\hbar \beta,I(\omega)]$ 
would allow for bridging a gap between information theoretical and pure thermodynamic results.
Work along this line will be reported soon. \\

\appendix

\begin{widetext}
\section{Reduced Classical Thermal State (Eq.~\ref{equ:equilstate})}
\label{app:ReducedClassState}
Consider the general expression for the physical situation described in Sec.~\ref{subsect:ThermalState}
\begin{equation}
\rho(\mathbf{p},\mathbf{q}) = \frac{1}{Z}
\exp 
	\left \{ -
	\left[ 
		H_{\mathrm{S}}(\mathbf{p}, \mathbf{q}) 
		+ \sum_{\mathfrak{j}}^{\mathfrak{N}} 
		\left[
		\frac{1}{2 m_{\mathfrak{j}}} \mathfrak{p}_{\mathfrak{j}} \cdot \mathfrak{p}_{\mathfrak{j}}
		+ \sum_{\mathfrak{i}}^{\mathfrak{N}} U_{\mathfrak{i,j}} 
			({\mathfrak{q}}_{\mathfrak{i}} - {\mathfrak{q}}_{\mathfrak{j}}) 
		+ \mathcal{V}_{\mathfrak{j}} ({\mathfrak{q}}_{\mathfrak{j}} - \mathbf{q})
	 	\right] 
	 \right] \beta 
	 \right \}.
\end{equation}
The reduced thermal state in Eq.~(\ref{equ:cgenequilstate}) reads
\begin{equation}
\rho_\mathrm{S}(\mathbf{p},\mathbf{q}) = 
\frac{1}{Z} \exp\left[- H_{\mathrm{S}}(\mathbf{p},\mathbf{q}) \beta \right]
\int \limits_{-\infty}^{\infty}
\prod_{\mathfrak{j}}^{\mathfrak{N}}\mathrm{d}\mathfrak{p}_{\mathfrak{j}} 
\mathrm{d}\mathfrak{q}_{\mathfrak{j}}
\exp 
	\left \{ -
		\sum_{\mathfrak{j}}^{\mathfrak{N}} 
		\left[
		\frac{1}{2 m_{\mathfrak{j}}} \mathfrak{p}_{\mathfrak{j}} \cdot \mathfrak{p}_{\mathfrak{j}}
		+ \sum_{\mathfrak{i}}^{\mathfrak{N}} U_{\mathfrak{i,j}} 
			({\mathfrak{q}}_{\mathfrak{i}} - {\mathfrak{q}}_{\mathfrak{j}}) 
		+ \mathcal{V}_{\mathfrak{j}} ({\mathfrak{q}}_{\mathfrak{j}} - \mathbf{q})
	 \right] \beta 
	 \right \},
\end{equation}
with 
\begin{equation}
Z = 
\int \limits_{-\infty}^{\infty} \mathrm{d} p \mathrm{d} q
\exp\left[- H_{\mathrm{S}}(\mathbf{p},\mathbf{q}) \beta \right]
\int \limits_{-\infty}^{\infty}
\prod_{\mathfrak{j}}^{\mathfrak{N}}\mathrm{d}\mathfrak{p}_{\mathfrak{j}} 
\mathrm{d}\mathfrak{q}_{\mathfrak{j}}
\exp 
	\left \{ -
		\sum_{\mathfrak{j}}^{\mathfrak{N}} 
		\left[
		\frac{1}{2 \mathfrak{m}_{\mathfrak{j}}} \mathfrak{p}_{\mathfrak{j}} \cdot \mathfrak{p}_{\mathfrak{j}}
		+ \sum_{\mathfrak{i}}^{\mathfrak{N}} U_{\mathfrak{i,j}} 
			({\mathfrak{q}}_{\mathfrak{i}} - {\mathfrak{q}}_{\mathfrak{j}}) 
		+ \mathcal{V}_{\mathfrak{j}} ({\mathfrak{q}}_{\mathfrak{j}} - \mathbf{q})
	 \right] \beta 
	 \right \}.
\end{equation}

After defining ${\mathfrak{q}}_{\mathfrak{k}} - \mathbf{q} = {\mathfrak{z}}_{\mathfrak{k}}$,
the integral over $\prod_{\mathfrak{j}}^{\mathfrak{N}} \mathrm{d}\mathfrak{p}_{\mathfrak{j}} 
\mathrm{d}\mathfrak{q}_{\mathfrak{j}}$ becomes
\begin{equation}
\int \limits_{-\infty}^{\infty}
\prod_{\mathfrak{j}}^{\mathfrak{N}} \mathrm{d}\mathfrak{p}_{\mathfrak{j}} 
\mathrm{d}\mathfrak{z}_{\mathfrak{j}}
\exp 
	\left \{ -
		\sum_{\mathfrak{j}}^{\mathfrak{N}} 
		\left[
		\frac{1}{2 \mathfrak{m}_{\mathfrak{j}}} \mathfrak{p}_{\mathfrak{j}} \cdot \mathfrak{p}_{\mathfrak{j}}
		+ \sum_{\mathfrak{i}}^{\mathfrak{N}} U_{\mathfrak{i,j}} 
			({\mathfrak{z}}_{\mathfrak{i}} - {\mathfrak{z}}_{\mathfrak{j}}) 
		+ \mathcal{V}_{\mathfrak{j}} ({\mathfrak{z}}_{\mathfrak{j}})
	 \right] \beta 
	 \right \}.
\end{equation}
So that,
\begin{equation*}
\rho_\mathrm{S}(\mathbf{p},\mathbf{q}) = 
\frac{
 \exp\left[- H_{\mathrm{S}}(\mathbf{p},\mathbf{q}) \beta \right]
\int \limits_{-\infty}^{\infty}
\prod_{\mathfrak{j}}^{\mathfrak{N}} \mathrm{d}\mathfrak{p}_{\mathfrak{j}} 
\exp 
	\left \{ -
		\sum_{\mathfrak{j}}^{\mathfrak{N}} 
		\left[
		\frac{1}{2 \mathfrak{m}_{\mathfrak{j}}} \mathfrak{p}_{\mathfrak{j}} \cdot \mathfrak{p}_{\mathfrak{j}}
	 \right] \beta 	 \right \}
\prod_{\mathfrak{j}}^{\mathfrak{N}} \mathrm{d}\mathfrak{z}_{\mathfrak{j}} 
\exp 
	\left \{ -
		\sum_{\mathfrak{j}}^{\mathfrak{N}} 
		\left[
			\sum_{\mathfrak{i}}^{\mathfrak{N}} U_{\mathfrak{i,j}} 
			({\mathfrak{z}}_{\mathfrak{i}} - {\mathfrak{z}}_{\mathfrak{j}}) 
		+ \mathcal{V}_{\mathfrak{j}} ({\mathfrak{z}}_{\mathfrak{j}})
	 \right] \beta 
	 \right \}}
{ \int \limits_{-\infty}^{\infty} \mathrm{d} p \mathrm{d} q 
\exp\left[- H_{\mathrm{S}}(\mathbf{p},\mathbf{q}) \beta \right]
\int \limits_{-\infty}^{\infty}
\prod_{\mathfrak{j}}^{\mathfrak{N}} \mathrm{d}\mathfrak{p}_{\mathfrak{j}} 
\exp 
	\left \{ -
		\sum_{\mathfrak{j}}^{\mathfrak{N}} 
		\left[
		\frac{1}{2 \mathfrak{m}_{\mathfrak{j}}} \mathfrak{p}_{\mathfrak{j}} \cdot \mathfrak{p}_{\mathfrak{j}}
	 \right] \beta 	 \right \}
\prod_{\mathfrak{j}}^{\mathfrak{N}} \mathrm{d}\mathfrak{z}_{\mathfrak{j}} 
\exp 
	\left \{ -
		\sum_{\mathfrak{j}}^{\mathfrak{N}} 
		\left[
			\sum_{\mathfrak{i}}^{\mathfrak{N}} U_{\mathfrak{i,j}} 
			({\mathfrak{z}}_{\mathfrak{i}} - {\mathfrak{z}}_{\mathfrak{j}}) 
		+ \mathcal{V}_{\mathfrak{j}} ({\mathfrak{z}}_{\mathfrak{j}})
	 \right] \beta 
	 \right \}
}.
\end{equation*}
Thus,
\begin{equation}
\rho_\mathrm{S}(\mathbf{p},\mathbf{q}) = 
\frac{
 \exp\left[- H_{\mathrm{S}}(\mathbf{p},\mathbf{q}) \beta \right]
}
{ \int \limits_{-\infty}^{\infty} \mathrm{d} p \mathrm{d} q 
\exp\left[- H_{\mathrm{S}}(\mathbf{p},\mathbf{q}) \beta \right]
} = Z_{\mathrm{S}}^{-1} \exp\left[- H_{\mathrm{S}}(\mathbf{p},\mathbf{q}) \beta \right],
\end{equation}
which corresponds to the canonical classical thermal state.

\section{Commutator $[ \hat{H}_{\mathrm{S}},\hat{V}]$ at fourth order}
\label{app:FourthOrder}
The equilibrium state at third order is given by

\begin{align}
\hat{\rho}=&e^{-\beta\hat{H}_{\mathrm{S}}}\left[ 1-\int\limits_{0}^{\hbar\beta}\mathrm{d}\sigma\,V_{\mathrm{SB}}(-i\hbar\sigma) \right. 
+ \left. \int\limits_{0}^{\hbar\beta}\mathrm{d}\sigma\int\limits_{0}^{\sigma}\mathrm{d}\theta\,\hat{V}_{\mathrm{SB}}(-i\hbar\sigma)\hat{V}_{\mathrm{SB}}(-i\hbar\theta)
-\int\limits_{0}^{\hbar\beta}\mathrm{d}\sigma\int\limits_{0}^{\sigma}\mathrm{d}\theta\int\limits_{0}^{\theta}\mathrm{d}\eta\hat{V}_{\mathrm{SB}}(-i\hbar\sigma)\hat{V}_{\mathrm{SB}}(-i\hbar\theta)\hat{V}_{\mathrm{SB}}(-i\hbar\eta) \right],
\end{align}
therefore
\begin{align}
\nonumber \left| \langle [ \hat{H}_{\mathrm{S}},\hat{V}_{\mathrm{SB}} ] \rangle \right| =& \left|\mathrm{tr}\left( [ \hat{H}_{\mathrm{S}},  \hat{S} ]\otimes \hat{B} e^{-\beta\hat{H}_{\mathrm{S}}}\left[ 1-\int\limits_{0}^{\hbar\beta}\mathrm{d}\sigma\,\hat{S}\otimes\hat{B}(-i\hbar\sigma) \right. \right. \right. + \left. \left. \left. \int\limits_{0}^{\hbar\beta}\mathrm{d}\sigma\int\limits_{0}^{\sigma}\mathrm{d}\theta\,\hat{S}\otimes\hat{B}(-i\hbar\sigma)\hat{S}\otimes\hat{B}(-i\hbar\theta) \right.  \right.\right. \\
&- \left. \left. \left. \int\limits_{0}^{\hbar\beta}\mathrm{d}\sigma\int\limits_{0}^{\sigma}\mathrm{d}\theta\,\int\limits_{0}^{\theta}\mathrm{d}\eta\,\hat{S}\otimes\hat{B}(-i\hbar\sigma)\hat{S}\otimes\hat{B}(-i\hbar\theta) \hat{S}\otimes\hat{B}(-i\hbar\eta) \right]  \right)\right|,
\end{align}
the first term that corresponds to  $\mathrm{tr}\left([ \hat{H}_{\mathrm{S}},  \hat{S} ]\otimes \hat{B} e^{-\beta\hat{H}_{\mathrm{S}}}\right)=\mathrm{tr}\left([e^{-\beta\hat{H}_{\mathrm{S}}}, \hat{H}_{\mathrm{S}}]  \hat{S} \otimes \hat{B} \right)=0$.
Therefore, after tracing out over the bath, we get that the lower bound in Eq.~(\ref{equ:thermodynbound}) is given by
\begin{align}
\nonumber |\langle[ \hat{H}_{\mathrm{S}},\hat{V}] \rangle| \propto &
\mathrm{tr}_{\mathrm{S}}
\left\{
[\hat{H}_{\mathrm{S}}, \hat{S}]\mathrm{e}^{-\hat{H}_{\mathrm{S}}\beta}
\left(
-\int_0^{\hbar \beta} \mathrm{d}\sigma \hat{S}(-\mathrm{i}\hbar\sigma) K(\sigma)
+ \int_0^{\hbar \beta} \mathrm{d}\sigma\int_0^{\sigma} \mathrm{d}\theta \hat{S}(-\mathrm{i}\hbar\sigma)\hat{S}(-\mathrm{i}\hbar\theta) L(\sigma,\theta) \right. \right. \\
&- \left. \left. \int_0^{\hbar \beta} \mathrm{d}\sigma\int_0^{\sigma} \mathrm{d}\theta\int_0^{\theta} \mathrm{d}\eta \hat{S}(-\mathrm{i}\hbar\sigma)\hat{S}(-\mathrm{i}\hbar\theta)\hat{S}(-\mathrm{i}\hbar\eta) M(\sigma,\theta,\eta)
\right)
\right\} \\
|\langle[ \hat{H}_{\mathrm{S}},\hat{V}] \rangle| \propto &
\mathrm{tr}_{\mathrm{S}}
\left\{
[\hat{H}_{\mathrm{S}}, \hat{S}]\mathrm{e}^{-\hat{H}_{\mathrm{S}}\beta}
\left(
\int_0^{\hbar \beta} \mathrm{d}\sigma \hat{S}(-\mathrm{i}\hbar\sigma) K(\sigma)
+ \int_0^{\hbar \beta} \mathrm{d}\sigma\int_0^{\sigma} \mathrm{d}\theta\int_0^{\theta} \mathrm{d}\eta \hat{S}(-\mathrm{i}\hbar\sigma)\hat{S}(-\mathrm{i}\hbar\theta)\hat{S}(-\mathrm{i}\hbar\eta) M(\sigma,\theta,\eta)
\right)
\right\} 
\end{align}
\end{widetext}
where
$\hbar K(\sigma) = \langle \hat{B}(-\mathrm{i}\sigma)\hat{B}(0) \rangle_{\mathrm{B}}$
denotes the two-time correlation of the bath operators given by  \cite{Ing02} and
$\hbar L(\sigma,\theta) = \langle \hat{B}(-\mathrm{i}\sigma)\hat{B}(-\mathrm{i}\theta)\hat{B}(0) \rangle_{\mathrm{B}}=0$ and $\hbar M(\sigma,\theta,\eta) = \langle \hat{B}(-\mathrm{i}\sigma)\hat{B}(-\mathrm{i}\theta)\hat{B}(-\mathrm{i}\eta)\hat{B}(0) \rangle_{\mathrm{B}}$
denotes the three and four-time correlation of the bath operators, respectively.

%

\begin{acknowledgments}
Discussions with Profs.~S. Mukamel and J. L. Garc\'ia Palacios and Dr. M. Campisi
are acknowledged with pleasure.
This work was supported by the \emph{Comit\'e para el Desarrollo de la Investigaci\'on}
--CODI-- of Universidad de Antioquia, Colombia under the Estrategia de Sostenibilidad, 
the grant number 2015-7631 and by the \emph{Departamento \
Administrativo de Ciencia, Tecnolog\'ia e Innovaci\'on} --COLCIENCIAS-- 
of Colombia under the grant number 111556934912, by NSERC and by the US Air Force 
Office of Scientific Research under contract numbers FA9550-13-1-0005 and
FA9550-17-1-0310.
\end{acknowledgments}

\bibliography{cntjcpv2}

\end{document}